# Quantum electric-dipole liquid on a triangular lattice

Shi-Peng Shen[1], Jia-Chuan Wu[2], Jun-Da Song[2], Xue-Feng Sun[2,3,4], Yi-Feng Yang[1], Yi-Sheng Chai[1], Da-Shan Shang[1], Shou-Guo Wang[1], James F. Scott[5] & Young Sun[1]

[1]Beijing National Laboratory for Condensed Matter Physics, Institute of Physics, Chinese Academy of Sciences, Beijing 100190, China
[2]Hefei National Laboratory for Physical Sciences at the Microscale, University of Science and Technology of China, Hefei 230026, China
[3]Key Laboratory of Strongly-Coupled Quantum Matter Physics, Chinese Academy of Sciences, Hefei, Anhui 230026, China
[4]Collaborative Innovation Center of Advanced Microstructures, Nanjing, Jiangsu 210093, China
[5]Cavendish Laboratory, University of Cambridge, J. J. Thomson Avenue, Cambridge, CB3 0HE, UK

**Geometric frustrations and quantum mechanical fluctuations may prohibit the formation of long-range ordering even at the lowest temperature, and therefore liquid-like ground states could be expected. A good example is the quantum spin liquid in frustrated magnets that represents an exotic phase of matter and is attracting enormous interests[1-3]. Geometric frustrations and quantum fluctuations can happen beyond magnetic systems. Here we propose that quantum electric-dipole liquids, analogs to quantum spin liquids, could emerge in frustrated dielectrics where antiferroelectrically coupled small electric dipoles reside on a triangular lattice. The quantum paraelectric hexaferrite $BaFe_{12}O_{19}$, in which small electric dipoles originated from the off-center displacement of $Fe^{3+}$ in the $FeO_5$ bipyramids constitute a two-dimensional triangular lattice[4], represents a promising candidate to generate the anticipated electric-dipole liquid. We present a series of experimental evidences, including dielectric permittivity, heat capacity, and thermal conductivity measured down to 66 mK, to reveal the existence of a nontrivial ground state in $BaFe_{12}O_{19}$, characterized by itinerant low-energy excitations with a small gap, to which we interpret as an exotic liquid-like quantum phase. The quantum electric-dipole liquids in frustrated dielectrics open up a fresh playground for fundamental physics and may find applications in quantum information and computation as well.**

Geometric frustrations arise on various triangle-based lattices like one-dimensional (1D) trestle lattice, two-dimensional (2D) triangular and kagome lattices, three-dimensional (3D) B-site spinel and pyrochlore lattices, and are typically investigated in spin systems[1-3]. It has become well known that the introduction of quantum fluctuations in geometrically frustrated magnets give rise to a rich variety of interesting quantum phases[5-8]. Especially, exotic quantum spin liquids (QSLs), characterized by either gapped or gapless itinerant excitations[8], have been theoretically predicted to show extremely intriguing phenomena. Compared with the impressive progress and diversity in theory, nevertheless, a clear identification of QSLs in real materials has proved challenging, with a very limited number of candidates reported so far[9-13].

Similar to the situation of spin lattices in magnets, geometric frustration can occur in lattices made of electric dipoles in dielectrics. In the case of small electric dipoles with significant quantum fluctuations persisted down to $T = 0$ K, exotic disordered quantum phases such as a quantum electric-dipole liquid (QEL), could be anticipated in certain conditions. In fact, some theoretical models proposed for ultracold dipolar particles trapped on 2D frustrated optical lattices have predicted topological quantum phases with fractional excitations[14]. In a QEL, the electric dipoles are highly entangled with one another in a form of quantum dimers (pairs of antiparallel dipoles) and continue to fluctuate in the resonating valence bond (RVB) state, a picture qualitatively similar to a QSL. However, we must emphasize that the QEL should have distinctive features from QSLs, because electric dipole and spin have important differences[15]. For instance, electric dipole neither has intrinsic angular momentum nor exhibits quantum precession as magnetic dipole (spin) does. Moreover, the nature of short-range and long-range interactions between electric dipoles is very different from that of spins[16]. This could lead to a very different phase diagram between QSLs and QELs.

Frustration in dielectrics has been previously studied in materials with competing ferroelectric (FE) and antiferroelectric (AFE) constituents such as the $KH_2PO_4$/$NH_4H_2PO_4$ (KDP-ADP) family or containing random-site impurities like $KTaO_3$:Li, which usually result in electric-dipole glasses similar to spin glasses[17,18]. However, geometric origin of frustrations and cooperative liquid-like quantum phases have been largely ignored in the studies of dielectrics. On another hand, the role of quantum fluctuations in dielectrics has been noticed since 1970s when people were studying the abnormal dielectric behavior of $SrTiO_3$ (ref. 19). It was proposed that quantum fluctuations in $SrTiO_3$ prevent the onset of long-range FE order so that a quantum paraelectric state persists down to zero temperature[20]. Since then, quantum paraelectricity has been reported in a number of perovskite oxides with similar structures to $SrTiO_3$, such as $CaTiO_3$, $EuTiO_3$, $KTaO_3$, *etc*. The quantum paraelectrics provide a new playground for the study of quantum critical phenomena[15], but it seems hopeless to search for the QELs in those perovskite quantum paraelectrics because their crystalline structures and FE interactions usually do not introduce geometric frustrations. In this Letter, we demonstrate that both geometric frustrations and quantum fluctuations can be simultaneously achieved in a unique frustrated dielectric ($BaFe_{12}O_{19}$). Our experiments measured down to 66 mK suggest that it has a very unusual liquid-like ground state, characterized by itinerant low-energy excitations with a small gap. We consider this nontrivial quantum phase as a possible candidate of QELs.

Recently we have discovered that the well-known M-type hexaferrites, such as $BaFe_{12}O_{19}$, belong to a completely new family of quantum paraelectrics[4]. Other hexaferrites containing the $FeO_5$ bipyramids in their crystal structures, such as the W-, Z-, X- and U-type hexaferrites, are also likely candidates of quantum paraelectrics[4]. The M-type hexaferrite $BaFe_{12}O_{19}$ is one of the most popular magnetic materials with a wide use in magnetic credit cards, bar codes, small motors, and low-loss microwave devices[21], due to its superior properties of ferrimagnetic ordering with a strong ferromagnetic moment and a very high Néel temperature (~ 720 K), high resistivity, as well as low cost of synthesis. The crystal structure of $BaFe_{12}O_{19}$ is shown in Fig. 1a. It can be described by a periodically stacking sequence of two basic building blocks – *S* block and *R* block along *c* axis. The $Fe^{3+}$ ions occupy three different kinds of sites: octahedral, tetrahedral, and bipyramidal sites. In particular, the $FeO_5$ bipyramids only exist in the middle of the *R*/*R*$^*$ blocks and form a triangular lattice in *ab* plane (Fig. 1b). Previous experiments including Mössbauer spectroscopy[22], x-ray diffraction[23], and neutron diffraction[24] have revealed the existence of off-equatorial displacements for $Fe^{3+}$ at Wyckoff position of 2b site inside the $FeO_5$ bipyramids to minimize the total energy, which results in two adjacent Wyckoff positions of 4e sites with a lowered symmetry (Fig. 1c). The off-equatorial displacement (4e-4e distances are 0.176(5) Å at 4.2 K and 0.369(5) Å at room temperature)[22] would induce a small local electric dipole *P* along *c* axis in each $FeO_5$ bipyramid (Fig. 1c). A dynamic displacement persists down to the lowest temperature due to the significant quantum tunneling between two 4e sites and the weak dipole-dipole coupling along *c* axis. Consequently, a quantum paraelectric behavior without long-range electric ordering has been observed in $BaFe_{12}O_{19}$ (ref. 4).

More importantly, these electric dipoles associated with the $FeO_5$ bipyramids reside on a triangular lattice in each *R*/*R*$^*$ block. Because the *R*/*R*$^*$ blocks are well separated by the *S*/*S*$^*$ blocks, this triangular lattice thus has a 2D feature. Consequently, a dielectric system with uniaxial (Ising-type) electric dipoles on a 2D triangular lattice is practically achieved in $BaFe_{12}O_{19}$ (Fig. 1c). If the neighboring dipole-dipole interaction favors anti-alignment, the system confronts frustrations and has a very large degeneracy of ground states. In this sense, $BaFe_{12}O_{19}$ would be a very unique quantum paraelectric other than those previously known, in which both geometric frustrations and strong quantum fluctuations may play an important role. Thus, the quantum paraelectric $BaFe_{12}O_{19}$ sets up a promising candidate to search the anticipated QELs, where an assembly of quantum dimers (pairs of dipoles) with either short-range or long-range entanglement continue to fluctuate (Fig. 1d). We then employ a series of experimental techniques to resolve the ground state of $BaFe_{12}O_{19}$.

A prerequisite of a QEL is the AFE interaction between neighboring dipoles. To confirm the AFE coupling in $BaFe_{12}O_{19}$, we have made a careful analysis on the low-temperature dielectric permittivity. As shown in Fig. 2a, the dielectric permittivity along $c$ axis ($\varepsilon_c$) of $BaFe_{12}O_{19}$ increases steadily with decreasing temperature but remains nearly constant below ~ 5.5 K. No dielectric phase transition is observed down to 1.5 K. This dielectric behavior evidences a quantum paraelectricity, similar to that in $SrTiO_3$. The quantum paraelectric behavior can be well described by the mean-field Barrett formula[25]:

$$\varepsilon = A + \frac{M}{(\frac{1}{2}T_1)\coth(\frac{T_1}{2T}) - T_0}$$ (1)

where $A$ is a constant, $T_0$ is proportional to the effective dipole-dipole coupling constant and the positive and negative values correspond to FE and AFE interactions, respectively. $T_1$ represents the tunneling integral and is a dividing temperature between the low temperature region where quantum fluctuation is important and the high temperature region where quantum effect is negligible. $M=n\mu^2/k_B$, where $n$ is the density of dipoles and $\mu$ denotes the local dipolar moment. After fitting the $\varepsilon_c$ below 160 K to the Barrett formula, we obtained $T_0 = -22.9(1)$ K and $T_1 = 47.3(1)$ K. The negative $T_0$ confirms the AFE coupling between electric dipoles. We note that recent first-principle calculations[28] also predicted the AFE interaction with frustration in $BaFe_{12}O_{19}$. The relative strength of quantum fluctuations can be estimated by ~ $|T_1/T_0| = 2.06$, which is likely high enough to favor a liquid ground state rather than an ordered or glass phase. The uniaxial anisotropy is evidenced by comparing the dielectric permittivity along $c$ axis with that in the $ab$-plane. As seen in the inset of Fig. 2a, the in-plane $\varepsilon$ decreases slowly with decreasing temperature (less than 1 for a temperature interval of 250 K). The absence of a paraelectric behavior in the $ab$-plane is consistent with the uniaxial electric dipoles along $c$ axis.

Further evidences of the AFE coupling in $BaFe_{12}O_{19}$ are presented in Fig. 2b. For those perovskite quantum paraelectrics with FE coupling, such as $SrTiO_3$, a moderated electric field is able to drive the quantum paraelectric state into a long-range ordered FE state. In strong contrast, for $BaFe_{12}O_{19}$, an external electric field of 5 kV/cm applied along c axis has no detectable influence on the dielectric permittivity. This inertness to external electric fields may indicate the AFE interaction in $BaFe_{12}O_{19}$. Moreover, the *P-E* loop at 2 K (the inset of Fig. 2b) shows a nearly linear response with quite small polarization up to a high electric field of 30 kV/cm, further implying the AFE coupling. It should be clarified that the magnetic moments of $Fe^{3+}$ at the bipyramidal sites are all parallel along $c$ axis in the $R/R^*$ blocks (see Supplementary Information) so that there are no magnetic frustrations on the triangular lattice[21,24,26].

The thermodynamic studies at temperatures as low as possible are crucial to identify the conjectured quantum liquid state, as they provide key informations on the spectrum of low-energy elementary excitations. Heat capacity and thermal transport measurements can probe the low-energy density of states as well as determine whether these low-energy excitations are localized or itinerant, and have been indispensably employed in the study of QSLs[27-29].

Since $BaFe_{12}O_{19}$ is a good insulator (see Supplementary Information) with long-range collinear ferrimagnetic ordering ($T_N$=720 K), both the electronic and magnon contributions to the thermal dynamics become negligible at very low temperatures[30]. Therefore, its thermodynamics at low enough temperatures should be dominated by the lattice contribution only, and the well-known $T^3$ dependence would be expected for both the heat capacity and thermal conductivity. Fig. 3a shows the heat capacity ($C_P$) of $BaFe_{12}O_{19}$ at low temperatures (T < 10 K). No sharp anomaly due to a phase transition could be detected down to 0.4 K, in accordance with the quantum paraelectric behavior. Unfortunately, the heat capacity data become scattered and noisy below ~ 1 K, possibly due to the very small values that reach the resolution limit of our equipment. Thus, a quantitative analysis of the heat capacity data is not possible.

The thermal conductivity provides more reliable and critical information on the low-lying elementary excitations, because it is sensitive exclusively to itinerant excitations and totally insensitive to localized entities that may cause the nuclear Schottky contribution and plague the heat capacity measurements at low temperatures[28,29]. For example, although the heat capacity study[27] suggested a gapless QSL in the frustrated triangular magnet $\kappa$-(BEDT-TTF)$_2$Cu$_2$(CN)$_3$, the thermal conductivity measurements[28] carried out down to 80 mK clarified instead a gapped QSL in the same material. We then have devoted a great effort to measure precisely the thermal conductivity of $BaFe_{12}O_{19}$ down to 66 mK.

Fig. 3b shows the thermal conductivity $\kappa$ measured in *ab*-plane as a function of temperature below ~ 1 K. $\kappa$ decreases rapidly with cooling, with a change more than 2 orders from 1 K to 100 mK. No anomaly due to a phase transition is observed down to 66 mK. The thermal conductivity divided by temperature as a function of $T^2$ is plotted in Fig. 3c. The data between 0.65 and 1 K exactly follow a linear relation with an extrapolation to the origin, in a good agreement with what expected for the phonon thermal conductivity, $\kappa = \beta T^3$, with $\beta$ = 0.098 WK$^{-4}$m$^{-1}$. Nevertheless, there is apparently an extra contribution below ~ 650 mK in addition to the normal phonon term, strongly suggesting the existence of abundant itinerant low-energy excitations other than phonons. Moreover, the thermal transport behavior at the zero temperature limit provides key information on the nature of these low-lying excitations. As seen in Fig. 3d, $\kappa/T$ in the $T \rightarrow$ 0 K limit tends to vanish rather than having a finite residual value, immediately implying the absence of gapless excitations. Instead, the data at the lowest temperature regime (T < 125 mK) can be fitted to

$$\kappa = \alpha \exp(-\Delta/k_B T) + \beta T^3 \quad . \tag{2}$$

The inset of Fig. 3 shows an Arrhenius plot of $\kappa^* = \kappa - \beta T^3$ in the lowest temperature region. The good linearity confirms the validity of Eq. (2). The best fit gives $\Delta$ = 0.16(1) K, which is much smaller than the effective dipole-dipole interaction constant $T_0$ (~ 22 K).

The exponential behavior of thermal conductivity at the zero temperature limit is very similar to that observed in the frustrated triangular magnet $\kappa$-(BEDT-TTF)$_2$Cu$_2$(CN)$_3$ where a QSL with gapped excitations ($\Delta = 0.46$ K) was identified[29]. Therefore, the thermal transport behavior excludes a frozen dipole glass or a classical paraelectric phase but strongly suggests an exotic liquid-like ground state.

In order to exclude the possibility that these itinerant low-lying excitations may have a magnetic origin, we further studied the influence of magnetic field on the thermal conductivity behavior. As seen in Fig. 3d, a high magnetic field of 14 T applied along the easy $c$ axis has no influence on the in-plane thermal conductivity in the lowest temperature range. The inertness of these low-lying excitations to external magnetic fields supports our argument that they stem from electric dipoles rather than spins.

Based on above experimental results, especially the thermal conductivity at the zero temperature limit, we conclude that an exotic ground state is likely established in the geometrically frustrated quantum paraelectric BaFe$_{12}$O$_{19}$. This liquid-like quantum phase is characterized by itinerant low-energy excitations with a small gap, which could be a candidate of anticipated QELs. As present transverse Ising models proposed for frustrated spin systems on a triangular lattice are inadequate for the frustrated electric dipoles with long-range interactions, a quantitative comparison between experiments and theories is not available at this stage. We expect that our experimental findings will stimulate theoretical efforts towards this subject. The present work serves as a start point of a fresh field and an abundance of exotic phenomena associated with geometrically frustrated quantum electric dipoles are awaiting ahead.

## Methods:

**Sample preparation.** The single crystal samples of BaFe$_{12}$O$_{19}$ were prepared by flux method and characterized with x-ray diffraction, as shown in Supplementary Fig. S1. Powders of BaCO$_3$, Fe$_2$O$_3$ and fluxing agent Na$_2$CO$_3$ were weighed to molar ratio of 1:1:1, then were mixed and well ground. The ground raw powder was put in Pt crucible and heated to 1250 °C for 24 h in the air, then cooled down to 1100 °C at a rate of 3 °C/min and finally quenched to room temperature.

**Dielectric measurements.** The dielectric measurements were carried out in a Cryogen-free Superconducting Magnet System (Oxford Instruments, TeslatronPT) down to 1.5 K. To measure the dielectric permittivity, silver paste was painted on the surfaces of a thin plate of crystal and annealed at 150 °C for about 30 mins to make good electrodes. An Agilent 4980A LCR meter was used to measure the dielectric permittivity with the frequency $f = 1$ MHz.

**Heat capacity and thermal transport measurements.** Heat capacity measurements were performed down to 0.4 K in a commercial Physical Properties Measurement System (PPMS, Quantum Design) using a $^3$He refrigerator. A thin-plate shaped sample with mass of 12.7 mg was used for this measurement. The contribution of attendant was measured separately and subtracted from the raw data. Thermal conductivity was measured between 60 mK and 1 K using the conventional steady-state "one heater, two thermometers" technique in a $^3$He-$^4$He dilution refrigerator (for details, see Zhao et al. *Phys. Rev. B* **91**, 134420 (2015)). A parallelepiped-shaped sample with size of 2.0 × 0.57 × 0.11 mm$^3$ was cut from the as-grown crystals for thermal conductivity measurements.

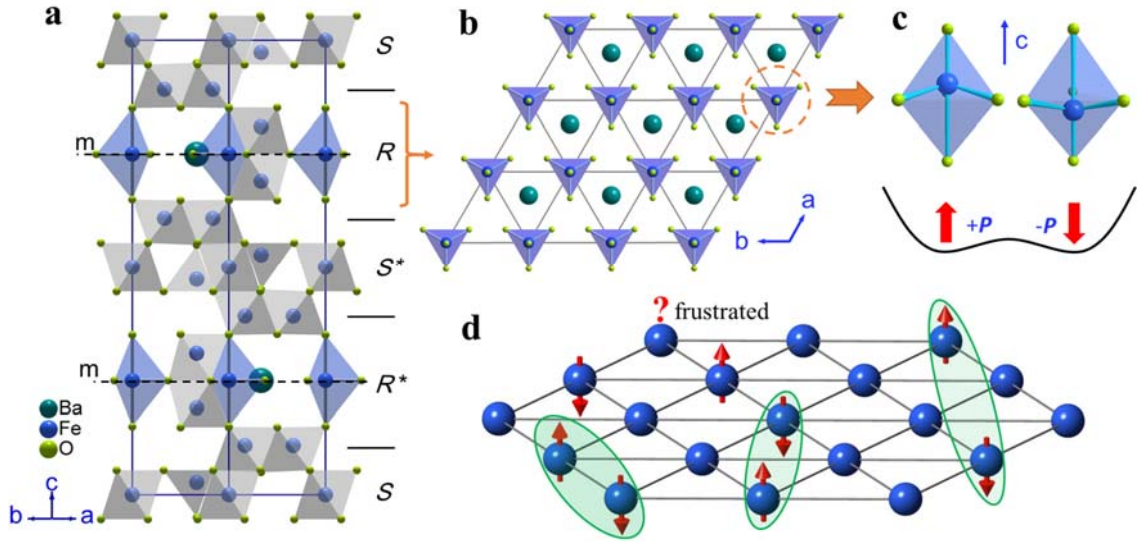

**Figure 1 Uniaxial electric dipoles on a triangular lattice in BaFe$_{12}$O$_{19}$.** (**a**) Crystal structure of the M-type hexaferrite BaFe$_{12}$O$_{19}$. It consists of alternate stacks of *S* and *R* blocks along *c* axis. The asterisk symbols indicate that the corresponding blocks rotate about *c* axis by 180°. The Fe$^{3+}$ ions occupy three different sites: octahedral, tetrahedral, and bipyramidal (blue) sites. A mirror plane (m) bisects equally the bipyramids in the *R/R*$^*$ blocks. (**b**) The 2D triangular lattice of FeO$_5$ bipyramids in each *R/R*$^*$ block. (**c**) Illustration of Fe$^{3+}$ off-equator displacements in the FeO$_5$ bypyramid. The upward or downward displacements at two 4e sites give rise to small electric dipoles along *c* axis. Quantum fluctuations between two 4e sites persist to *T*=0 K. (**d**) Frustrated electric dipoles on a triangular lattice. Each site contains an Ising-type electric dipole while the neighboring interactions favor anti-alignment. Quantum dimers with either short-range or long-range entanglement continue to fluctuate and results in a QEL.

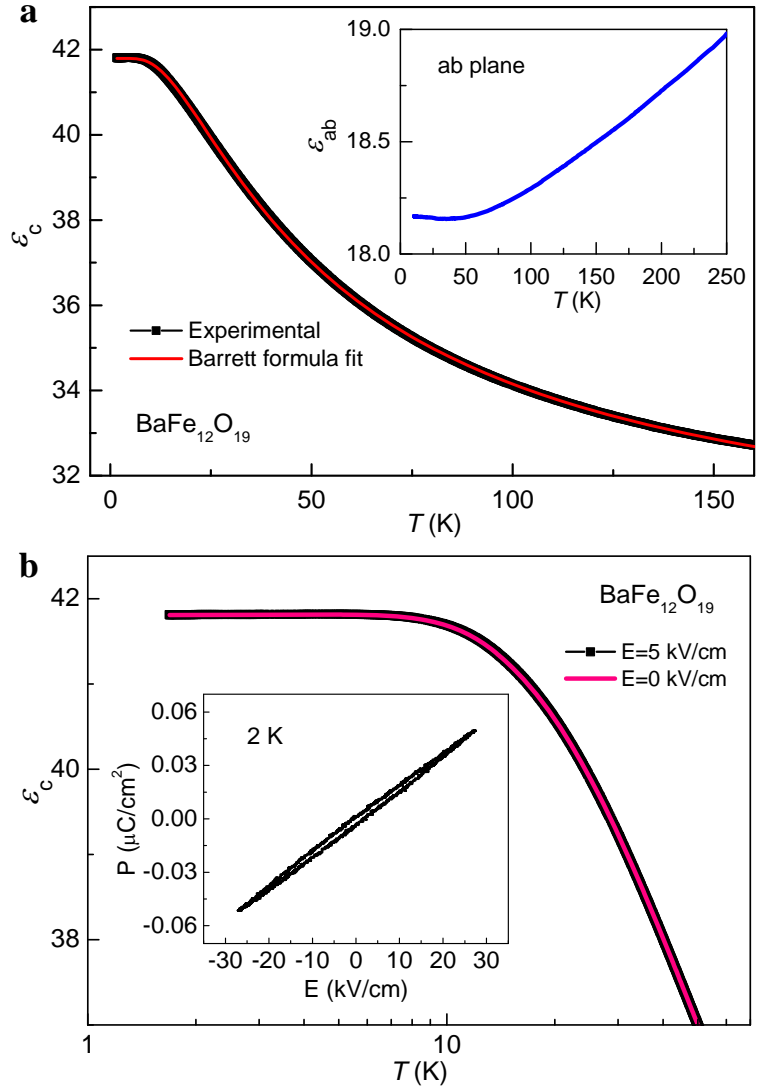

**Figure 2 Dielectric permittivity of BaFe$_{12}$O$_{19}$.** (a) The temperature dependence of *c*-axis dielectric permittivity $\varepsilon_c$. The red solid line is the fitting curve to the Barrett formula. The negative fitting parameter $T_0 = -22.9(1)$ K suggests the AFE interaction. The inset shows the dielectric permittivity measured along the [100] direction in *ab*-plane. (b) The temperature dependence of dielectric permittivity $\varepsilon_c$ measured with DC bias electric fields. A bias electric field of 5 kV/cm has no detectable influence on the quantum paraelectric behavior. The inset shows the *P-E* loop at 2 K. The nearly linear response and the low values of *P* imply the AFE coupling.

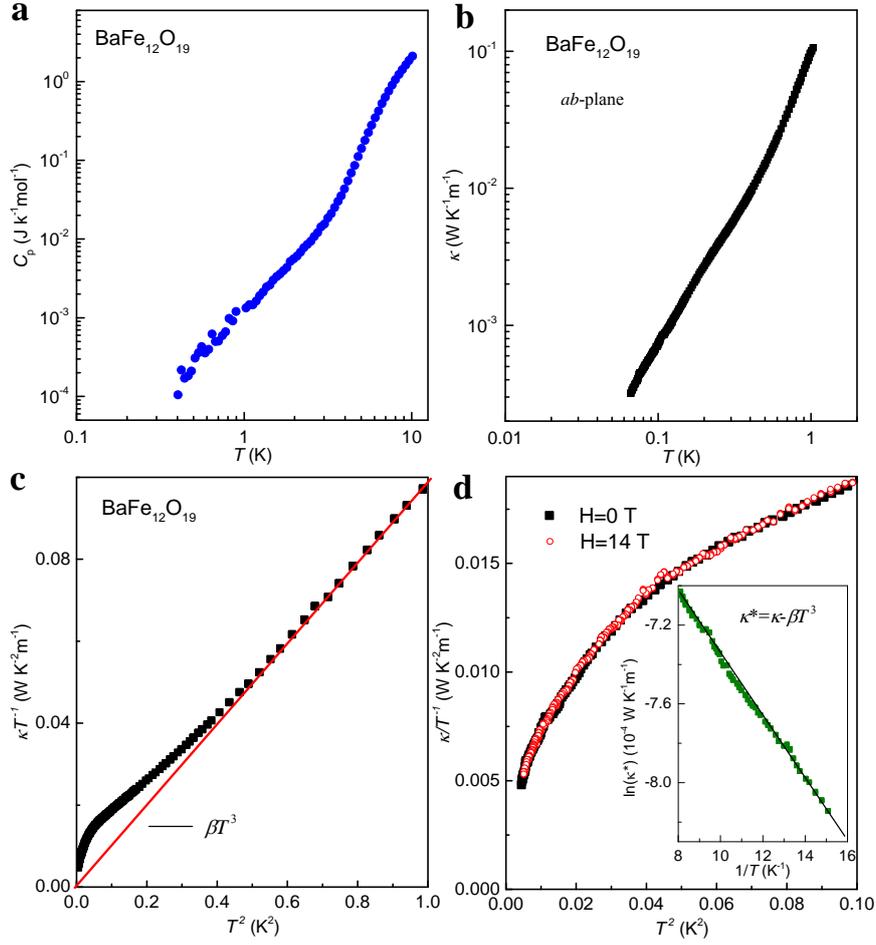

**Figure 3 Heat capacity and thermal conductivity of BaFe$_{12}$O$_{19}$.** (**a**) Heat capacity $C_p$ as a function of temperature. No phase transition is detected down to 0.4 K. (**b**) Thermal conductivity $\kappa$ measured in *ab*-plane as a function of temperature. No anomaly due to a phase transition is observed down to 66 mK. (**c**) The *ab*-plane thermal conductivity divided by temperature plotted as a function of $T^2$ below ~ 1 K. The red solid line represents the expected thermal conductivity of phonons, $\kappa = \beta T^3$, with $\beta = 0.098$ WK$^{-4}$m$^{-1}$. Apparently, there is an extra contribution beside the phonon thermal conductivity below ~ 650 mK. (**d**) The $\kappa T^{-1}$ versus $T^2$ plot in the lowest temperature region. $\kappa/T$ tends to vanish at the $T \to 0$ K limit. An applied magnetic field of 14 T along *c* axis has no influence on the *ab*-plane thermal conductivity in this low-temperature region. The inset shows the Arrhenius plot of $\kappa^* = \kappa - \beta T^3$ below ~ 125 mK. The good linearity suggests gapped excitations with a small gap ~ 0.16 K.

# Supplementary Information

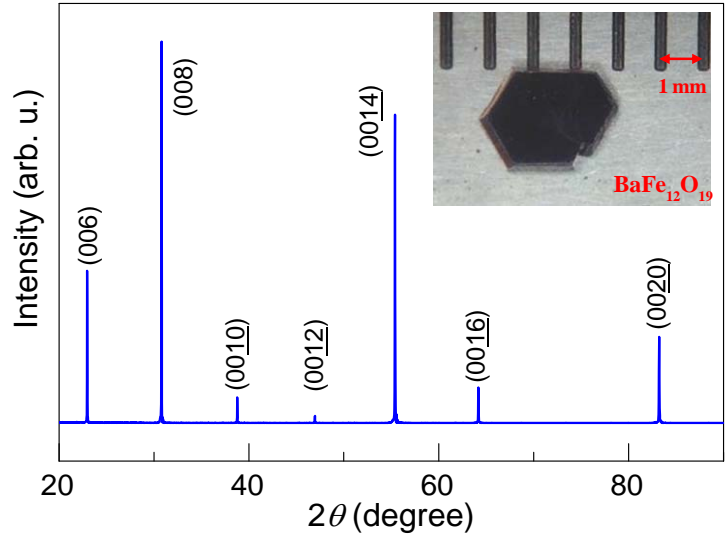

**Fig. S1** The single-crystal x-ray diffraction pattern of BaFe$_{12}$O$_{19}$ at room temperature. The inset shows a picture of flux-grown single crystals.

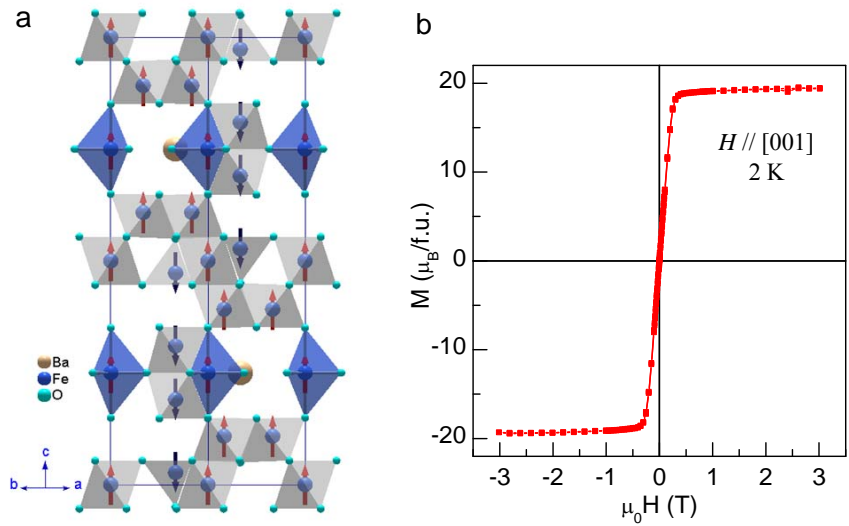

**Fig. S2** (a) Magnetic structure of BaFe$_{12}$O$_{19}$. The magnetic moments at the FeO$_5$ bipyramidal sites are parallel to each other so that there is no magnetic frustration on the triangular lattice. (b) The *M-H* hysteresis curve of BaFe$_{12}$O$_{19}$ measured at 2 K along *c* axis. It is consistent with a long-range collinear ferrimagnetic ordering.

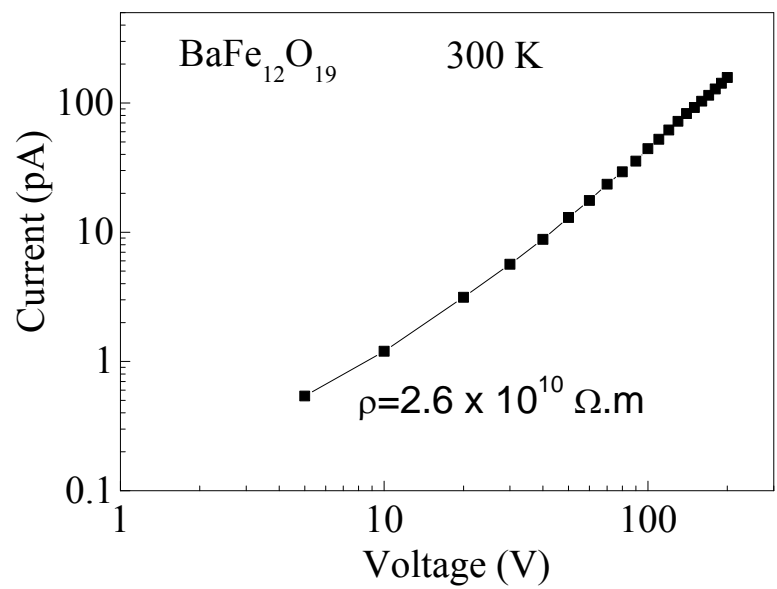

**Fig. S3** I-V characteristic and resistivity along c axis of $BaFe_{12}O_{19}$ measured at room temperature. The sample is highly insulating even at room temperature, suggesting a good quality of the grown single crystals.